# Bilayer graphene nanoribbons junction with aligned holes exhibiting high *ZT* values

Shuo Deng[†‡], Lijie Li[*‡] and Paul Rees[*‡]
[†]Wuhan University of Technology, Wuhan 430070, China
[‡]College of Engineering, Swansea University, Swansea SA1 8EN, UK
[*]Emails: L.Li@swansea.ac.uk; P.Rees@swansea.ac.uk

**Abstract**: We investigate the thermoelectric performance of armchair graphene nanoribbon (AGNR), bilayer GNRs junction (BGNRJ) and BGNRJ with holes (BGNRJ-H) by the first principles calculation with non-equilibrium Green function. It is found that the BGNRJ-H exhibits high *ZT* values of 9.65 and 5.55 at 300K. The reason of these significantly larger *ZT* values than previously observed has been calculated due to reduced thermal conductivity and enhanced electrical conductivity. The low thermal conductance comes from the van der Waals (vdW) interaction between two graphene layers. The increased electrical conductivity can be attributed to the coupling effect of aligned holes in both layers. It is found from analysis results that the electron transmission of the BGNRJ-H is much stronger than a normal BGNRJ, which gives rise to the higher electrical conductance and outstanding *ZT* values.

**Keywords**: Graphene nanoribbon junction with aligned holes; Thermoelectric; First principles method

## 1. Introduction

Thermoelectric materials have been promising in recycling heat energy. However, it is well known that current thermoelectric technologies are severely hindered by low conversion efficiency [1]. The energy conversion efficiency of thermoelectric materials is characterized by the figure of merit $ZT = S^2 G_e T/\kappa$, where the *S*, $G_e$, *T* and $\kappa$ are Seebeck coefficient, electrical conductance, temperature and thermal conductance, respectively [2]. The standard method to improve *ZT* is to increase Seebeck coefficient, electrical conductance and to reduce thermal conductance. However, it is difficult to improve *ZT* by simply increasing electrical conductance or reducing thermal conductance because there is a strong coupling between the electron and phonon transport [3, 4]. Hence the key to have a higher *ZT* value is to find a trade-off between the Seebeck coefficient, electrical and thermal conductance. A review of strategies used to improve the thermopower and reduce the thermal conductivity was given in reference[3].

Graphene, the first 2-dimensional (2D) material[5], shows many outstanding properties, such as high carrier mobility[6, 7], superior mechanical properties[8, 9] and high thermal conductivity[10, 11]. During the last two years, bilayer graphene structures with strong interlayer coupling and high temperature superconductivity have been experimentally investigated[12-14]. The thermoelectric performance of graphene has been studied, but the figure of merit was unsatisfactory (*ZT*~0.05[15]) because of the high thermal conductivity[10]. Therefore, the specially designed nanostructures have been realized to suppress the thermal conductivity without greatly reducing electrical conductance [16-21]. In these nanostructures, graphene nanoribbons (GNRs) have demonstrated higher figure of merit thanks to the finite size effect[22-28]. In 2009, Ouyang *et al* reported that the Seebeck coefficient of GNRs is higher than the pristine graphene because of the edge effect [29]. After three years, Jin *et al* explored that the armchair GNR (AGNR) exhibited a high figure of merit at 300 K ($ZT\sim6$)[22]. One of earlier explorations on the thermoelectric performance of bilayer GNRs junction (BGNRJ) was reported by Nguyen *et al*[30]. They found that the thermal conductivity of BGNRJ is much weaker than monolayer GNRs due to the vdW interaction between two layers. The BGNRJ structure efficiently limits the thermal conductivity in the normal direction to the 2D plane, leading to a new approach to improve the thermoelectric performance [31, 32]. Although the thermoelectric transport in graphene nanoribbons and derivatives such as defected single layer graphene was extensively studied in recent years, the structure of bilayer graphene nanoribbons junction with an aligned hole is first proposed in our manuscript.

In this paper, we explore the electron transport and thermoelectric performance of BGNRJs with two holes (BGNRJ-H) aligned together in the top and bottom of the junction. The previous experiments demonstrated that bilayer graphene nanoribbons and graphene defects were produced by self-assembly of molecules [33] and electron beam irradiation at high current density within a nanoscale region [34], respectively. Therefore, synthesis of bilayer graphene nanoribbons junction with aligned holes is feasible. First principles method and Boltzmann transport theory are employed, which have been widely applied in prior research on standard BGNRJs[31, 32]. The predicted *ZT* value of the new BGNRJ-H structure can be significantly improved to 9.65 and 5.55 at $\pm 0.61$ eV chemical potentials at 300K.

## 2. Computational procedure

The electronic properties of BGNRJ-H are calculated by the first-principles method implemented in the Quantum Atomistix ToolKit (ATK2018) simulation tools[35]. A 3.35Å spacing between the two nanoribbons is obtained after using the semi-empirical corrections by the Grimme DFT-D2 model to take account the long-range vdW interaction[36]. The atomic positions are fully relaxed until the magnitude of force on each atom becomes less than 0.01 eV/Å. The generalized gradient approximation (GGA) with the Perdew-Burke-Ernzerhof functional (PBE), cut-off energy of 150 Ry and 10×10×1 k-points grid were used. Particularly for the bandgap calculation, we use the HSE exchange–correlation functional to achieve accurate results. We calculated band structure and orbital projected density of state (PDOS) of the three graphene nanoribbons. In the Figs. 1(a) and 1(b), the bandgap of AGNR and BGNRJ are 1.47 eV and 0.72 eV, respectively. However, for the BGNRJ-H, the bandgap closing is shown in Fig. 1(c). Moreover, the PDOS for each graphene nanoribbon shows that the band has the contribution mainly from the C-*p* orbital and C-*d* orbital.

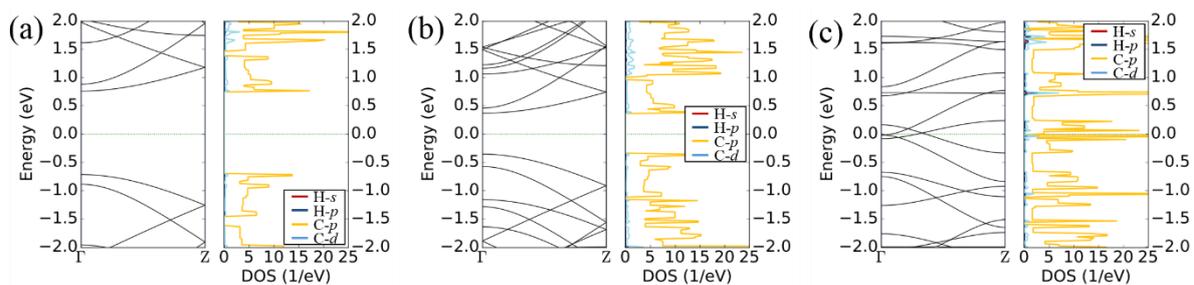

*Fig. 1. The band structure and orbital projected density of state of AGNR (a), BGNRJ (b) and BGNRJ-H (c).*

In our transmission calculation, the Brillouin zone of the device is sampled by a 1×1×200 *k*-mesh and a double-zeta polarized for all atoms. We adopt an enough vacuum spacing of at least 25 Å in the normal direction to the electron transport plane to avoid interaction of the periodic boundary conditions. The electron and phonon transport properties are calculated by the non-equilibrium Green function method. The total electronic transmission function $T_e(E)$ is calculated as

$$T_e(E) = Tr[t^+ t](E) \qquad (1)$$

From the Landauer-Buttiker equation, the current across the device can be calculated as

$$I = \frac{e}{h} \int \left( f_L(E) - f_R(E) \right) T_e(E) dE \qquad (2)$$

where, $f_L(E)$ and $f_R(E)$ are the Fermi distribution functions of the left and right electrode, respectively. The functional $L_n(T)$ is defined as

$$L_n(T) = \frac{2}{h} \int_{-\infty}^{+\infty} (E - E_F)^n T_e(E) \left( -\frac{\partial f_{FD}(E, E_F)}{\partial E} \right) dE \qquad (3)$$

$f_{FD}(E, E_F)$ is the Fermi-Dirac distribution function. Accordingly, the electron conductance ($G_e = e^2 L_0/h$), the Seebeck coefficient ($S = L_1/eTL_0$) and heat transport coefficient of electrons ($\kappa_e = (L_0 L_2 - L_1^2)/htL_0$) can be calculated. From the phonon transmission function $T_p(\omega)$, the heat transport coefficient of phonon ($\kappa_p(T)$) can be obtained from

$$\kappa_p(T) = \frac{1}{2\pi} \int_0^\infty \hbar w \, T_{ph}(\omega) \frac{\partial f_{BE}(\omega, T)}{\partial T} d\omega \qquad (4)$$

where, $f_{BE}(\omega, T)$ is the Bose-Einstein distribution function.

## 3. Results and discussion

Fig. 2(a) shows the details of the device models. In order to compare different junction structures, we show the schematic diagrams of BGNRJ and BGNRJ-H devices, respectively. In the BGNRJ-H model, six carbon atoms are removed from the center of top and bottom junction. The length and width of the overlap region in the BGNRJ is 7 unit-cells along the transport direction. The 7 unit-cells width is the minimum width required because 6 carbon atoms are removed from the center of the junction in the BGNRJ-H model. In this work, the variations taken into consideration are the junction structures and the width of the GNRs. The electronic transmission spectra of BGNRJ and BGNRJ-H are given in Fig. 2(b). The transmission peaks of BGNRJ and BGNRJ-H are asymmetrical about the Fermi level (0 eV). The transmission peak values of the first conduction channel for BGNRJ and BGNRJ-H are about 0.25 and 1.77 for both the *p*-type doping (E<0) and *n*-type doping (E>0). The electron transportation through BGNRJ-H is much stronger than BGNRJ. There is no electron transmission from -0.60 eV to 0.60 eV for both cases as the AGNRs exhibit semiconducting properties. Fig. 2(c) shows the phonon transmission spectra of BGNRJ and BGNRJ-H. It can be seen clearly that the transmission values of BGNRJ are slightly higher than the BGNRJ-H between 0 eV and 0.19 eV.

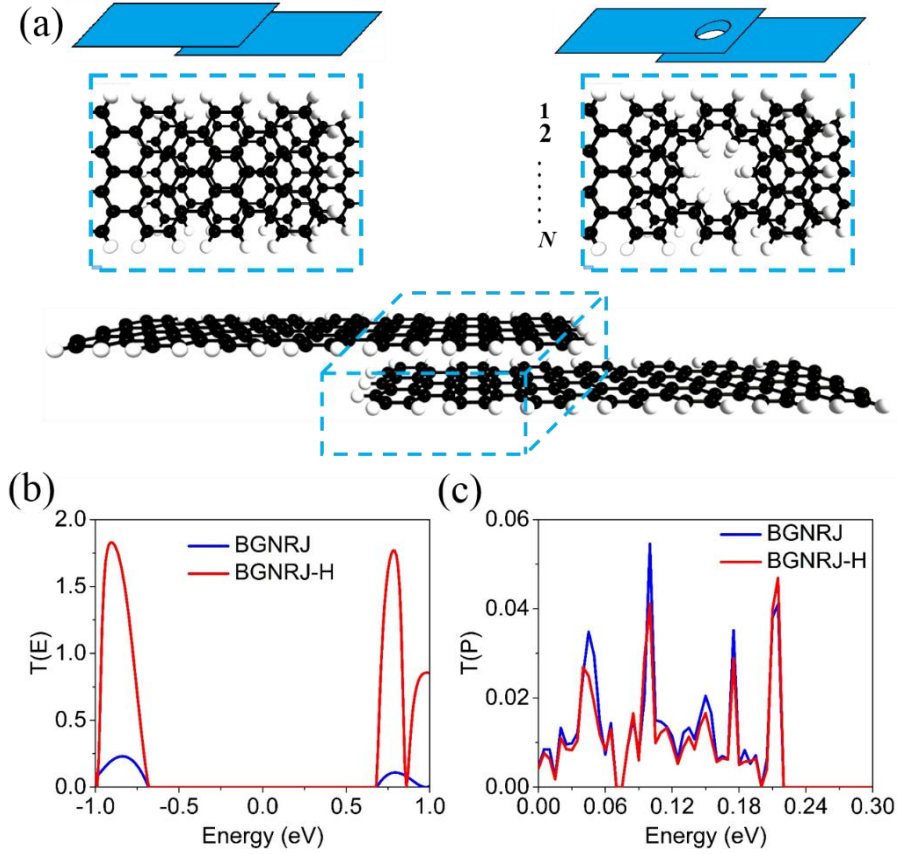

*Fig. 2. Schematic diagram (a), electronic transmission spectra (b) and phonon transmission spectra (c) of BGNRJ and BGNRJ-H.*

Fig. 3(a) show the calculated *I-V* and *dI/dV* curves for AGNRs, BGNRJ and BGNRJ-H with the same length and width. Between 0 - 1 *V* bias voltage range, the AGNR has the largest current in three structures ($I_{AGNR}$> $I_{BGNRJ-H}$> $I_{BGNRJ}$). When the bias voltage is smaller than 0.5 *V*, the current displays an increasing trend with the increase of the bias voltage. However, after 0.5 *V*, the current tends to be stable, even slightly reduces with the increase of bias voltage, which is obviously in the *dI/dV* curves (Fig. 3(b)) at 0.6 V and 0.9 V. Similar results were shown in prior research[37]. By integrating the calculated electron and phonon transmission spectra, the electrical conductance, thermal conductance and Seebeck coefficient of AGNR, BGNRJ and BGNRJ-H can be readily obtained at 300 K. As shown in Fig. 3(c)-2(f), the difference between the Fermi energy ($E_F$) and the DFT-predicted Fermi energy ($E_F^{DFT}$) indicates the chemical potential. In the Fig. 3(c), the electronic structures of the junction and monolayer sections of the devices are different, which results in the valleys and peaks in the electrical conductance curves. The maximum of electrical conductance in AGNR is found to be 76 µA, which is about 9.5 times and 1.6 times higher than the BGNRJ and GGNRJ-H, respectively. Fig. 3(c) shows a trend of $G_{e-AGNR}$> $G_{e-BGNRJ-H}$> $G_{e-BGNRJ}$, which is exactly matching with the results from the *I-V* curves that $I_{AGNR}$ > $I_{BGNRJ-H}$ > $I_{BGNRJ}$. The developments of thermal conductance of AGNR, BGNRJ and BGNRJ-H (Fig. 3(d)) have a similar trend as the electrical conductance. Compared with the thermal conductance of the AGNR, there is a strong reduction for BGNRJ and BGNRJ-H because vdW interaction hinders the heat transport between two layers AGNR[30, 38]. The Seebeck coefficients of three devices are shown in Fig. 3(e), the maximum value is about 2.1 mV/K, which is around 21 times higher than the pristine graphene (~0.1 mV/K)[39]. However, the Seebeck coefficients of AGNR, BGNRJ and BGNRJ-H are very close, which means that the electrical conductance and thermal conductance are two key contributing parameters to lead to a promising figure of merit for the BGNRJ and BGNRJ-H. The *ZT* values of AGNR, BGNRJ and BGNRJ-H as shown in the Fig. 3(f). It is clear that four high *ZT* values are more than 1 for the BGNRJ and BGNRJ-H at room temperature. The highest *ZT* is 9.65, which is better than most previously reported figures of merit at 300 K[16, 32, 40].

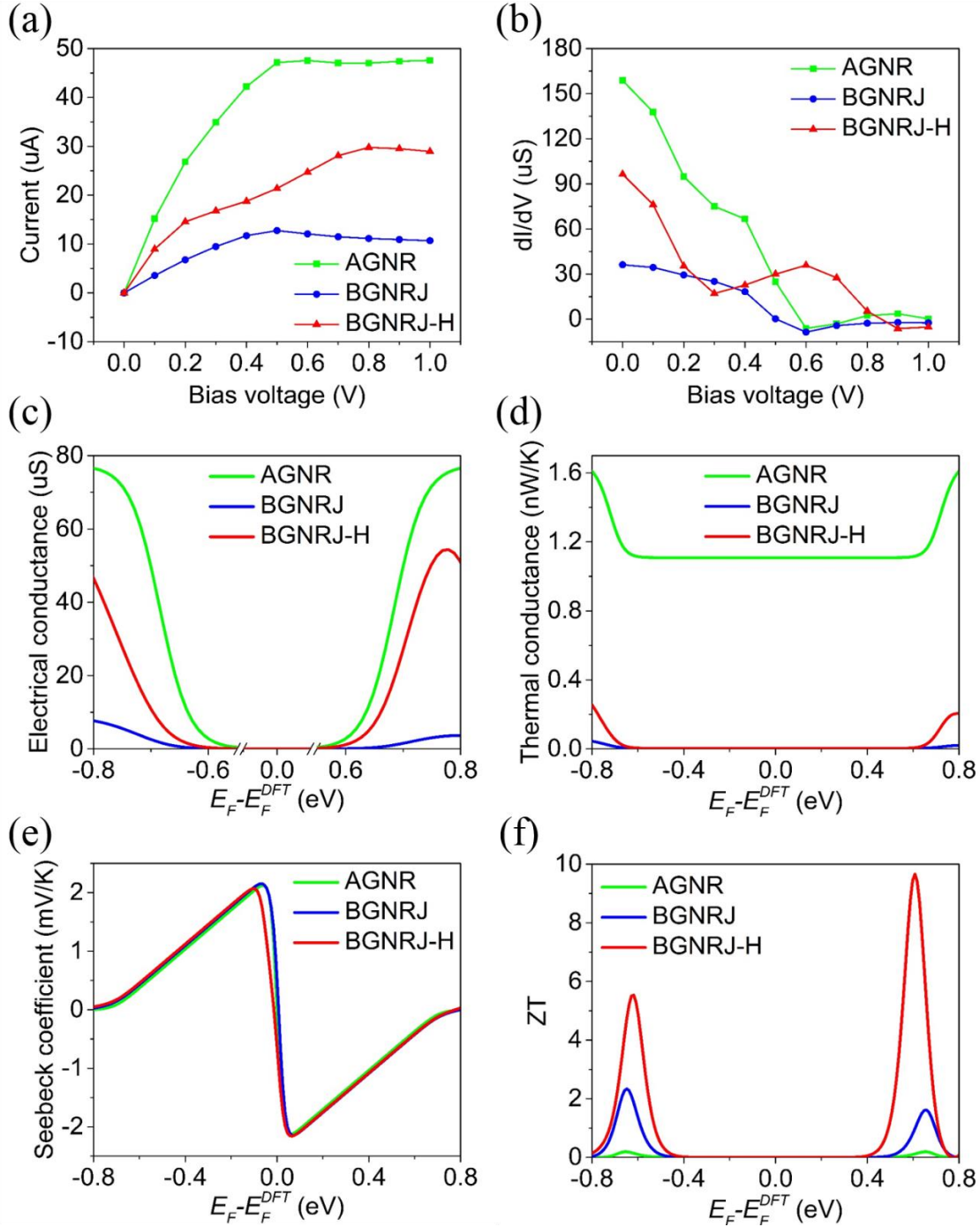

*Fig. 3. I-V (a), dI/dV (b), electrical conductance (c), thermal conductance (d), Seebeck coefficient (e) and ZT values (f) of AGNR, BGNRJ and BGNRJ-H.*

We calculated phonon band structure of BGNRJ-H in the Fig. 4. The BGNRJ-H is stable because no negative frequencies in the Brillouin zone of Γ-Z-T-Γ. Moreover, we examine the stability of the constructed BGNRJ-H by Born-Oppenheimer molecular dynamics (BOMD) simulation in ATK. In the NVT Nose Hoover scheme, we set temperature at 450 K and the simulation has run through 3 ps thermostat timescale and 1 fs time step. After the BOMD simulation, only small changes appear in the BGNR-J structure, which demonstrates a good thermal stability for BGNRJ-H.

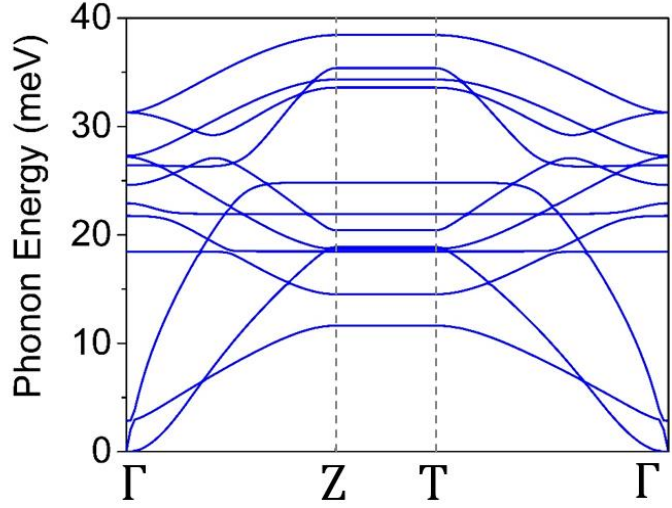

*Fig. 4. The phonon band structure of BGNRJ-H.*

In order to explore the reason for the high *ZT* values in the BGNRJ and BGNRJ-H, we have conducted thermal conductance and electron transport calculations. Fig. 5(a) and 5(b) show calculated thermal conductance from contributions of phonons and electrons in the chemical potential range around the high *ZT* peaks. It is shown that the electron contribution to the overall thermal conductance is less than the phonon contribution for BGNRJ around the high *ZT* values. However, for the BGNRJ-H, the phonon contribution to thermal conductance is comparable to that of electrons in the range of $\pm 0.61$ eV chemical potentials. To investigate the effect of temperature on the thermoelectric performance of BGNRJ-H in the 0.61 eV, we calculated the variation of $k_p$, $k_p$ and $k$ in Fig. 5(c). The phonon contribution to the whole thermal conductance is more than the electron contribution when T<300 K. However, the electron thermal conductance is higher than the phonon thermal conductance after T>300 K. The whole thermal conductance increases with the rise of temperature. As shown in Fig. 5(d), the *ZT* values of BGNRJ-H increases to the maximum when T<400 K, while the *ZT* reduces at high temperature (T>400 K). To explain the reason of non-monotonicity for *ZT*-temperature function in BGNRJ-H, we calculate the variation of $S^2GeT$ in Fig. 5(d). Obviously, the value of $S^2GeT$ increases monotonically with the temperature, which means the transition of *ZT* value at 400 K is induced by the variation of electron thermal conductivity.

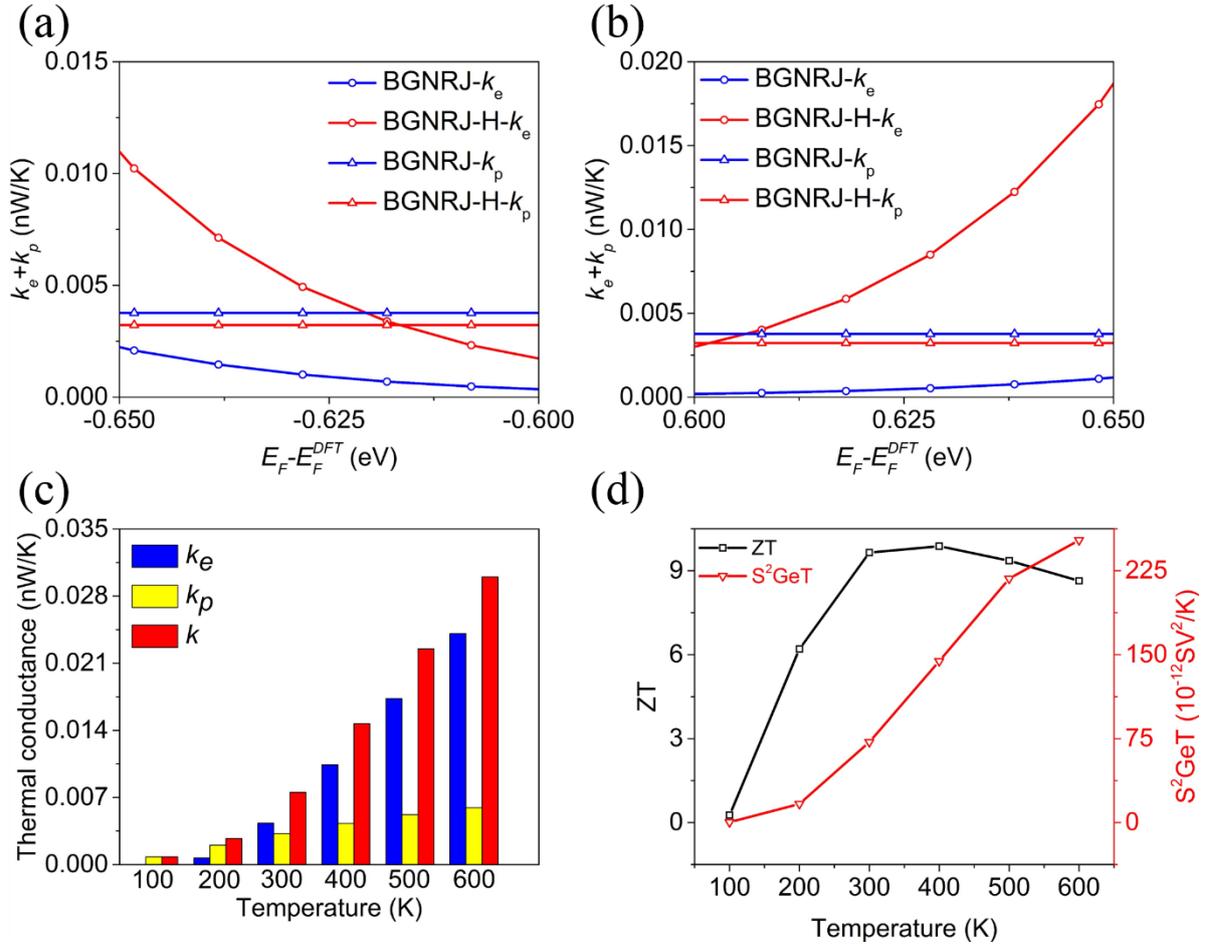

*Fig. 5. Thermal conductance of phonons and electrons parts (a)-(b), variation of $k_p$, $k_p$ $k$, ZT and $S^2GeT$ with temperature (c)-(d).*

The total thermal conductance of BGNRJ-H is slightly higher than the BGNRJ (see Table 1), which means that the electron transport plays a key role for high *ZT* values of BGNRJ-H. As shown in the Table 1, the electron conductance of BGNRJ-H is 1.66 μS and 0.82 μS, which is around 4.3 times and 1.3 times higher than the BGNRJ (0.38 μS and 0.62 μS) at the *ZT* peaks. Fig. 6(a) and (b) show the electron transmission pathways of BGNRJ and BGNRJ-H devices in the central region, which graphically demonstrate the intensity and path of how electrons flow through atoms. The current contains two different electron transport channels: C-C bond current (electron flows through the C-C bond) and C-C hop current (electron hops between two non-bonded C atoms of the same cell). It is shown that the C-C bond current is stronger than the C-C hop current, and the electron transports in the BGNRJ and BGNRJ-H are mainly along the armchair edge. Moreover, from the color bar of the transmission pathways, the total electron transmission coefficient of BGNRJ-H is much higher than the BGNRJ, which is matching with the results from the *I-V* and electrical conductance results. Fig. 6(c) and 6(d) show the cut-plane representation of transmission eigenstate through the BGNRJ and BGNRJ-H devices central regions. The transmission eigenstates correspond to a scattering state coming from the left electrode and traveling towards the right. The eigenstate will always have a small amplitude in the right part of the device. Compared with the BGNRJ (Fig. 6(c)), the eigenstates from BGNRJ-H (Fig. 6(d)) have much larger amplitudes in the right part of the device, which indicates that a higher transmission probability of an incoming electron traveling through the central region and into the right electrode. The transmission eigenstates in the left part of the scattering region are composed of an incoming (right moving) and a reflected (left moving) state, which leads to interference effects. In the Fig. 6(c), the transmission eigenstates of BGNRJ have a very small amplitude from the middle part to the right end of the device, this is attributed to the fact that a small portion of the scattering eigenstate is transmitted and majority of them are reflected at the junction. However, after creating aligned holes in both the top and bottom layers of junction (BGNRJ-H), the scattering eigenstate reflected at the junction becomes weaker and much of them have transmitted to the right part of the device. This much enhanced electron coupling between two aligned holes in the graphene nanoribbon junction is the most innovative discovery, which underpins the record high values of *ZT*. If only making a vacancy or bi-vacancy around the center of BGNRJ, it will not affect the electron transmission too much as a vacancy or bi-vacancy is too small compared with a hole. This also coincides with the results in the Fig. 7, where the hole in the wider BGNRJs (*N*=15, *N*=23) has relatively little effect on the electron transmission.

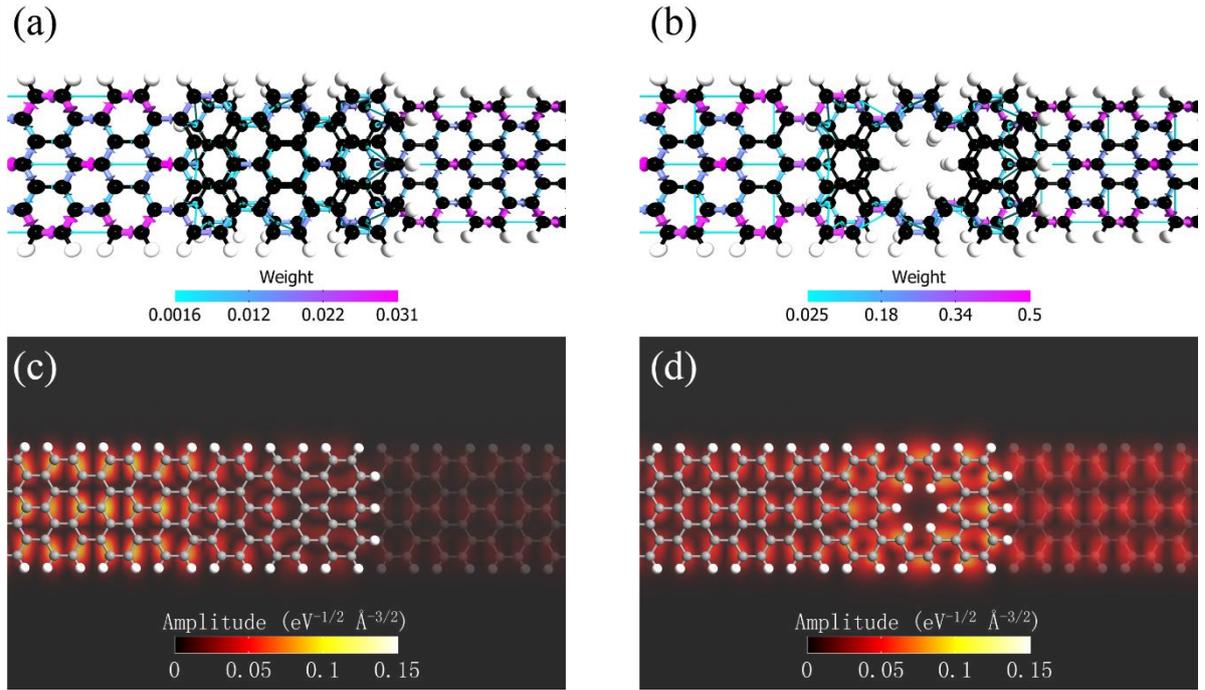

*Fig. 6. Electron transmission pathways (a)-(b) and electron transmission eigenstate (c)-(d) of BGNRJ and BGNRJ-H.*

With the calculated transport coefficients of electrons and phonons, we can evaluate the figure of merit of BGNRJ and BGNRJ-H. As shown in Table I, there is very little differences on the Seebeck coefficient and overall thermal conductance of BGNRJ and BGNRJ-H, which implies that electron conductance should play a key role for the outstanding *ZT* values of BGNRJ-H. In the transmission eigenstates analysis, a good transmission has been demonstrated, indicating macroscopically a good impedance matching thank to the aligned holes on both graphene layers.

Moreover, in the Fig. 3(f), all the *ZT* curves are nearly symmetric around the 0 eV. Hence, we can further enhance *ZT* by applying appropriate *p*-type or *n*-type doping to the BGNRJ-H. For the BGNRJ-H, there are two peak values (~ 9.65, ~ 5.55) at around ±0.61 eV, which means *n*-type doping are more favorable than *p*-type doping to increase the figure of merit of BGNRJ-H. We compared our results with prior research about thermoelectric performance of graphene nanostructures at 300 K in Table II. In previous results, the highest *ZT* value about 6.11 can be obtained in the twisted bilayer GNR junction. Hence, the aligned holes bilayer GNR junction is a better candidate for the thermoelectric devices.

Table I. Calculated 300 K *ZT* values and corresponding electronic and photonic transport coefficient for AGNR, BGNRJ and BGNRJ-H.

| Structure | $E_F - E_F^{DFT}$ (eV) | S (mV/K) | G (μS) | $\kappa_e$ (nW/K) | $\kappa_p$ (nW/K) | ZT |
|---|---|---|---|---|---|---|
| AGNR | 0.65 | 0.21 | 15.86 | 0.039 | 1.11 | 0.19 |
|  | -0.65 | 0.21 | 15.86 | 0.039 | 1.11 | 0.19 |
| BGNRJ | 0.65 | 0.26 | 0.38 | 0.0012 | 0.0038 | 1.61 |
|  | -0.65 | 0.27 | 0.62 | 0.0022 | 0.0038 | 1.61 |
| BGNRJ-H | 0.61 | 0.38 | 1.66 | 0.0043 | 0.0032 | 9.65 |
|  | -0.61 | 0.39 | 0.82 | 0.0046 | 0.0032 | 5.55 |

Table II. *ZT* values and chemical potential of different graphene nanostructures

| Structure | ZT | $E_F - E_F^{DFT}$ (eV) |
|---|---|---|
| Monolayer GNR[29] | 0.11 | 0.29 |
| Mixed GNR[41] | 1.12 | 0.81 |
| Kinked GNR[15] | 0.59 | 1.22 |
| Defected GNR[42] | 3.91 | 0.60 |
| Nanopore GNR[43] | 5.20 | -0.32 |
| Multilayer GNR Junction[30] | 1.81 | 0.55 |
| Twisted Bilayer GNR Junction[32] | 6.11 | 0.68 |
| Single Molecule GNR Junction[44] | 0.42 | -0.05 |
| Quantum Dots Junction[45] | 1.28 | 0.79 |
| Aligned Holes Bilayer GNR Junction (our work) | 9.65 | 0.61 |

The electrical conductance, thermal conductance, Seebeck coefficient and *ZT* values of BGNRJ-Hs with different widths have been calculated and shown in Figs. 7(a)-(d). Here we consider three widths, i.e. *N*=7, 15 and 23, which were chosen as the hole will be exactly in the center for these *N* numbers. In Figs 7(a) and (b), there is a sharp increase of electrical conductance and thermal conductance at both sides of Fermi level. Fig. 7(c) shows the Seebeck coefficient calculated for different widths at 300 K. As it is clearly seen that *N*=7 has the maximum Seebeck coefficient of 2.1 mV/K. The maximum Seebeck coefficient of *N*=7 is 2.8 times higher than that of *N*=15 and 10.2 times higher than *N*=23. We note that the Seebeck coefficient reduce with the increase of the nanoribbon width, which is similar to what was reported in the prior works [22, 46]. The calculation results of *ZT* values at 300 K are shown in Fig. 7(d). The *ZT* values of *N*=7 is nearly 9.8 times higher than *N*=15 and 23. These observations suggest that by choosing a smaller width, one can obtain a higher Seebeck coefficient with much improved thermoelectric performance. More importantly, recent experiments demonstrated precise synthesis at atomic level of AGNRs-7 by self-assembly of molecules, which boosts up the feasibility of this theoretical work [47, 48].

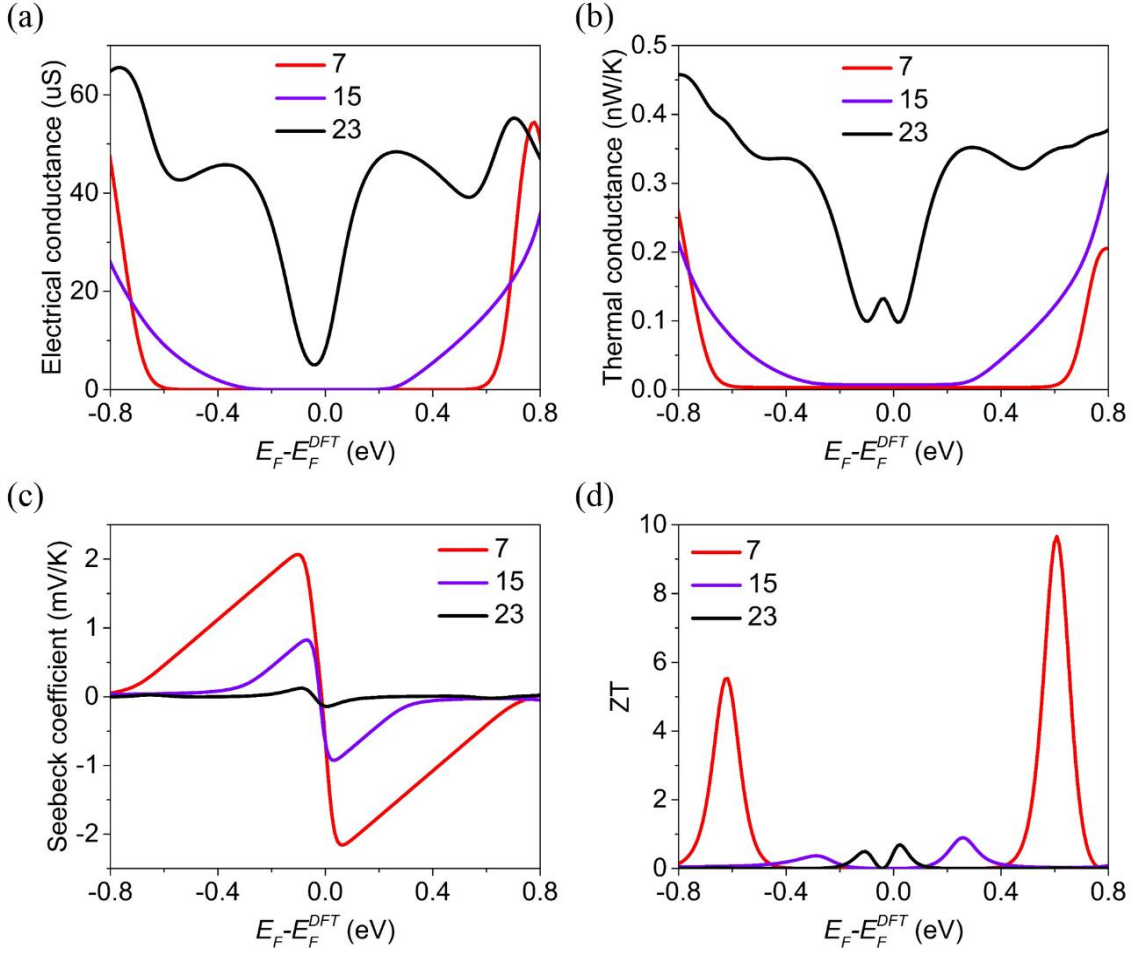

*Fig. 7. Electrical conductance (a), thermal conductance (b), Seebeck coefficient (c) and ZT values (d) of BGNRJ-H with N=7, 15 and 23.*

## 4. Conclusion

In summary, we have investigated the electron and phonon transportations of AGNR, BGNRJ and BGNRJ-H using the first principles calculation. Ultra-high *ZT* values of 9.65 and 5.55 have been achieved at around $\pm 0.61$ eV for the BGNRJ-H at 300K. It is interpreted that the reason of very high *ZT* values of BGNRJ-H is due to its reduced thermal conductivity and significantly enhanced electrical conductivity at junction with aligned holes. The low thermal conductance comes from the vdW interaction between two layers. Compared with the BGNRJ, the reflection effect of the scattering electron transmission eigenstate at the junction becomes weaker and a good transmission has been demonstrated after creating aligned holes in both top and bottom layers of the junction (BGNRJ-H). The high *ZT* values of BGNRJ-H makes it a compelling candidate device for thermoelectric applications.

**Acknowledgement**

This work was supported by the China Scholarship Council (CSC) and the European Regional Development Fund (ERDF) for the funding of the Solar Photovoltaic Academic Research Consortium (SPARC II).